# 2D PZT MEMS Resonant Scanner Using a Three-Mask Process

Mehrdad Khodapanahandeh[a,#], Parviz Zolfaghari[b,#,*], Hakan Urey[b,*]

[a] *Department of Biomedical Science and Engineering, Koç University, Istanbul, 34450, Turkey*

[b] *Department of Electrical Engineering, Koç University, Istanbul, 34450, Turkey*

[#] *Authors contributed equally*

[*] *Corresponding authors*

mkhodapanahandeh21@ku.edu.tr, pzolfaghari@ku.edu.tr, hurey@ku.edu.tr

**Abstract**

This work presents the design, simulation, fabrication, and characterization of a novel architectural compact two-dimensional (2D) resonant MEMS scanning mirror actuated by thin-film lead zirconate titanate (PZT). The device employs an innovative mechanically coupled dual-axis architecture fabricated using a three-mask process on an SOI-PZT deposited wafer, significantly reducing system complexity while achieving high performance. The scanner integrates a 1 × 1.4 mm oval mirror within a 7 × 4.7 mm die, actuated by PZT thin-film elements optimized for resonant operation at 3.6 kHz (vertical) and 54.2 kHz (horizontal) under 12 Vp-p periodic pulse driving. The system achieves optical scan angles of 4.8° and 11.5° in vertical and horizontal directions, respectively, with quality factors of 750 (vertical) and 1050 (horizontal). These values contribute to high scanning bandwidth-efficiency products of 24.2 deg·mm·kHz (vertical) and 623 deg·mm·kHz (horizontal), among the higher values reported for 2D PZT-MEMS scanners. Finite element analysis confirmed minimal stress and mirror deformation, and experimental validation demonstrated excellent agreement with simulation results. This architecture demonstrates the feasibility of high-resolution laser scanning, as required in applications such as OCT, LiDAR, and displays, by achieving performance levels in line with those used in such systems.

## 1. Introduction

Microelectromechanical systems (MEMS) mirrors are an important component of numerous advanced applications, including light detection and ranging (LiDAR) [1-11], augmented reality (AR) and virtual reality (VR) displays [12, 13], ophthalmological imaging, optical coherence tomography (OCT) [14-16], and compact projectors in head-up displays like smart glasses [17-20]. In most of the laser scanning systems, two MEMS mirrors are coordinated to scan in a two-dimensional scanning [4, 21, 22]. The horizontal mirror typically operates at kilohertz frequencies, while the vertical mirror functions at lower frequencies, often in the hundreds of hertz range [4, 22]. The mirrors are designed at different scales to minimize diffraction effects: horizontal mirrors typically measure around 1 mm, whereas vertical mirrors are larger, often spanning several millimeters [4, 23].

MEMS scanners are highly adaptable and have their applications in the broadest of fields, from display and laser technology to biomedical imaging and laser printing. There has been a growing demand for compact, high-performance laser scanners, especially in the context of mobile pico-projector applications. In such applications, actuators must achieve high operating frequencies, large scan angles, and low driving voltages, all while preserving a compact footprint. Conventional scanning systems typically depend on either electromagnetic or electrostatic actuation. Electromagnetic actuators require bulky off-chip magnets, which contribute to an increased device size, whereas electrostatic actuators, despite their compact form, generally demand high driving voltages [24-29]. In contrast, piezoelectric actuation [9-11, 30-33], particularly using thin-film lead zirconate titanate (PZT), presents a compelling alternative, offering significant angular deflection at low voltages within a compact form factor.

Thanks to its high-power density and small footprint, recent developments in piezoelectric MEMS have shown the efficiency of thin-film PZT in a range of applications, including ultrasonic motors, transducers, and micromirrors. Besides, PZT actuators have a unique advantage in closed-loop control by facilitating direct deflection measurements [34]. These measurements are obtained through piezoelectrically generated voltage or current, thereby eliminating the need for external sensing components. Despite these advantages, several technical challenges remain, particularly in achieving high-frequency operation. Significant design limitations persist due to damping losses and the restricted deflection resulting from the inherent stiffness of PZT beams [35-37].

Designing a high-performance 2D scanner for raster scanning remains challenging; however, several bi-resonant piezoelectric dual-axis scanners have been successfully demonstrated. In 2012, Chen et al. presented a 2D Lissajous scanner actuated by an external PZT element [35]. The device exhibited resonance frequencies of 25 kHz and 560 Hz along the two axes, enabling image generation at a resolution of 640×480. Gu-Stoppel et al. proposed an alternative gimbal-free design with three degrees of freedom, implemented using four PZT-coated cantilevers [38]. This configuration supported torsional resonance at 28.6 kHz and a piston mode at 21 kHz. Tani et al. developed a scanner incorporating a meander spring structure capable of operating in both resonant and quasi-static modes [39]. In one implementation, both axes were actuated using the meander spring configuration, achieving non-resonant optical scan angles of ±17.2° and ±11.2° for the outer and inner axes, respectively. Additionally, a hybrid scanning system was implemented, integrating a slow-moving outer axis based on a meander spring with a faster torsional inner axis actuated by two unimorph beams [40]. The device successfully demonstrated 2D scanning: the fast horizontal axis operated in resonance at 11.2 kHz with a 39° optical angle, while the slow vertical axis functioned at 60 Hz, achieving a ±7.25° optical range under a 40 Vp-p drive voltage. This work marks the first documented demonstration of a piezoelectrically actuated raster scanner for 2D scanning [22].

To explore a potential pathway forward, this study presents an innovative architectural design that integrates both scanning axes into a single MEMS device, resulting in substantial size reduction and improved operational efficiency. The proposed 2D MEMS scanner features a compact oval micromirror (1 × 1.4 mm), actuated by PZT elements previously developed by our research group [41, 42]. The 1 mm dimension of the mirror is employed along the resonant axis, while the entire device is encapsulated within a 7 × 4.7 mm frame, resulting in a compact and high-performance scanning system. Nevertheless, each design enhancement introduces inherent trade-offs that must be carefully managed to meet performance requirements, particularly the need to balance the fast-axis and slow-axis scanning frequencies, which directly influence the resolution and frame rate [4]. In the presenting work, the resonance frequency was increased from 40 kHz, as reported in previous studies for the 1D version [42], to approximately 54 kHz in the 2D configuration for the fast-axis, while preserving the 1 mm mirror dimension. Although the proposed prototype 2D device demonstrates enhanced resonance frequencies, it exhibits a reduced scanning angle compared to our previous 1D designs [42]. The remainder of this paper is organized as follows: Section 2 describes the device design and fabrication process; Section 3 presents the finite-element model; Section 4 discusses the experimental characterization and results; Section 5 provides a discussion; and Section 6 presents the concluding remarks.

## 2. Device Architecture and Fabrication
### A. Scanner Design

The proposed novel compact 2D PZT MEMS scanning mirror architecture consists of three cascaded frames, as illustrated in the fabricated optical image in Figure 1(a). The outer frame functions as a support structure, carrying the other two frames and enabling vertical scanning. This frame is anchored to the substrate by two torsional beams. Four PZT cantilever actuators, also anchored to the substrate, are connected to the outer frame via folded beams. The middle frame, shaped like four wings, functions as the horizontal actuator and is connected to the outer frame. The inner frame, which holds the scanning mirror, is specifically designed to minimize dynamic deformation and preserve mirror integrity. This frame is connected to the middle frame through torsional flexures. This frame is connected to the middle frame via torsional flexures. For actuation, PZT thin-film actuators are integrated on each wing of the middle frame and on the cantilevers for bidirectional actuation.

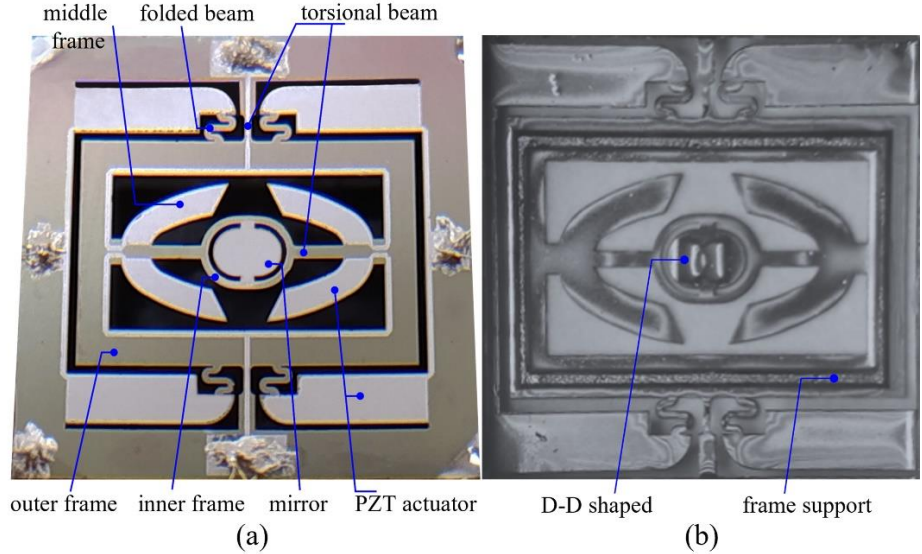

*Figure 1. (a) Top view of the fabricated compact 2D PZT MEMS scanning mirror wire-bonded to a printed circuit board (PCB). The major structural components of the device are labeled. The PZT cantilevers drive the outer frame in the vertical direction, while the wing-shaped PZT actuators on the middle frame act as complementary drive electrodes, causing the inner frame and mirror to resonate horizontally. The inner frame is designed to mechanically isolate the mirror from the torsional flexure, minimizing deformation. (b) The bottom view of the scanner shows the dynamic support structures (D–D shaped).*

As shown in Figure 2, the scanning electron microscope (SEM) image provides a detailed view of the chip dimensions. The die measures 1 cm × 1 cm, with the movable components occupying an area of 8 mm × 8 mm. The overall device thickness is 130 μm (Figure 2b). The outer frame is rectangular, measuring 4 mm × 7.5 mm. Torsional beams, which anchor the outer frame to the substrate alongside folded beams, are 130 μm wide (Figure 2c). The middle frame is connected to the outer frame via primary suspension flexures, each measuring 220 μm in length and 520 μm in width. The inner frame is linked to the middle frame by two torsional flexures, each 480 μm long and 270 μm wide. To reduce peak stress concentrations, the flexure anchor points are rounded with fillets. Within the inner frame, the scanner incorporates an oval mirror measuring 1 mm × 1.4 mm.

Two principal design strategies are implemented to minimize dynamic deformation in both the mirror and outer frame. First, a 175 μm-thick D–D-shaped reinforcement rim is retained from the silicon handling layer beneath the mirror and outer frame. This rim crosses the rear side of the mirror and outlines a narrower edge of the outer frame, thereby enhancing structural stiffness, as illustrated in Figure 1(b). Second, the mirror is mechanically decoupled from the torsional beams via the inner frame.

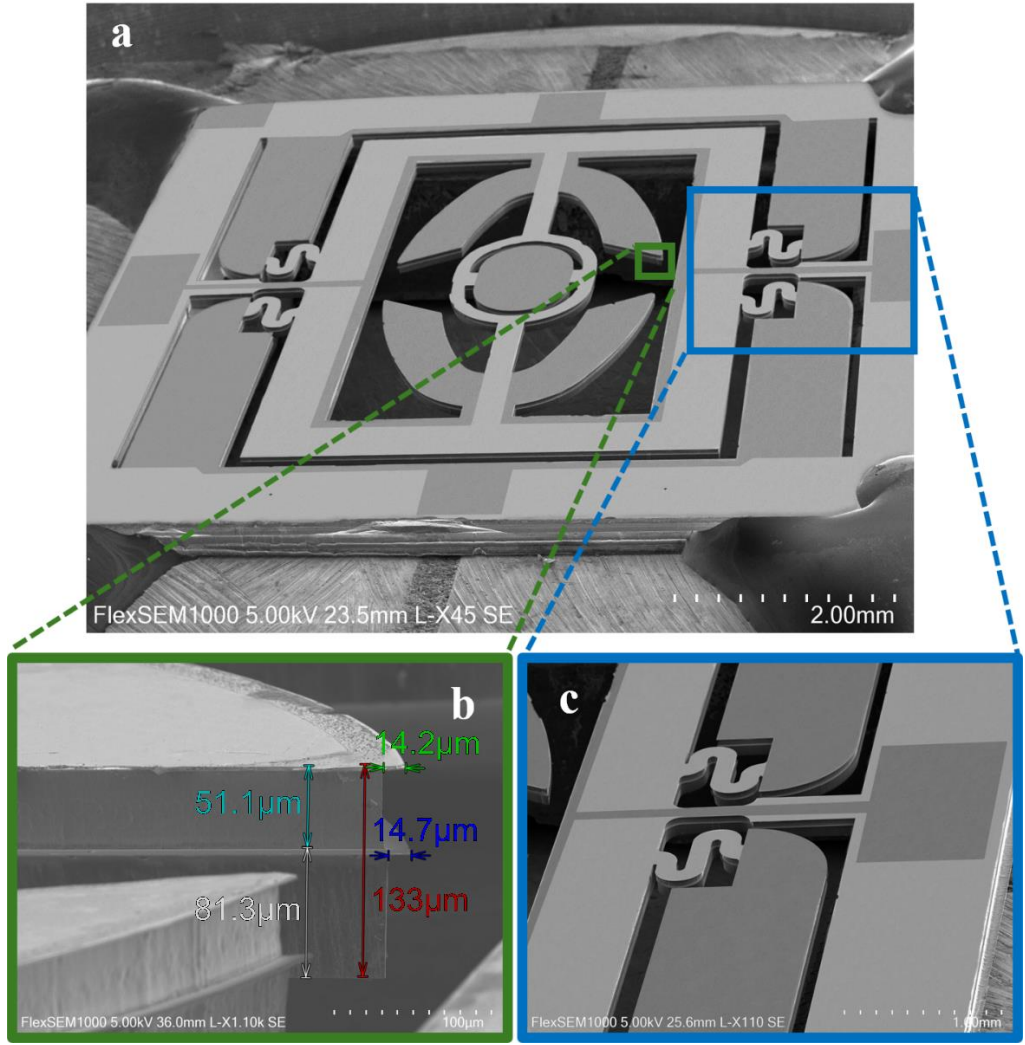

*Figure 2. (a) SEM image of the entire die. (b) Cross-sectional view of the released device. Although deep reactive ion etching (DRIE) was performed vertically at 90°, over-etching is visible in this image. The device consists of a 50 μm device layer and an 80 μm handle layer on a silicon-on-insulator (SOI) wafer, resulting in a total thickness of 130 μm. An over-etch depth of 14 μm is observed in this SEM image. (c) SEM image showing the folded beam and torsional flexure mechanism of the outer frame.*

## B. Fabrication

In this work, the employed microfabrication process relies on a three-mask technique, which is derived and modified from the six-mask process developed within our research group previously by Baran et al. [41, 42]. A silicon-on-insulator (SOI) wafer featuring a 50 µm-thick device layer, a 2 µm-thick buried oxide (BOX) layer, and a 375 µm-thick handle substrate was used in this study based on availability in our laboratory. Both sides of the wafer were uniformly deposited with a 1.5 µm thickness of silicon dioxide ($SiO_2$). A 30 nm titanium (Ti) adhesion layer followed by a 100 nm platinum (Pt) layer was deposited on the device layer side, serving as the bottom electrode. A 1.5 µm Ba(Barium)-doped PZT thin film was then deposited by sputtering onto the top surface of the device layer. The wafer was fabricated by PEIMAC Corporation and then diced into 1.2 cm × 1.2 cm chips. The process began by spin-coating a 2 µm-thick layer of AZ 5214E photoresist onto the chip at 4000 rpm for 100 seconds, followed by a soft bake at 110 °C for 55 seconds. The first mask was patterned directly onto the chip using the Heidelberg MLA 100 maskless aligner. The exposed regions were then developed in AZ 726 MIF to define the top electrodes, interconnect lines, mirror, and wire-bonding pads. To eliminate residual nanoscale photoresist, the chip was subjected to an $O_2$ plasma treatment, preparing the surface for subsequent deposition. A 30 nm chromium (Cr) adhesion layer and a 170 nm aluminum (Al) layer were then deposited using an E-Beam & Thermal Evaporator (KJL PVD 200 Pro Line) at a deposition rate of 0.5 Å/s, as shown in Figure 3(b).

Following metal deposition, the chip was immersed in acetone to complete the lift-off process. The chip was then spin-coated with photoresist, the same process as in the first lithography step, and patterned with the second mask, which is the final top-side lithography mask of the device. The PZT layer was etched using a wet chemical solution with a composition of 30 mL deionized (DI) water, 70 mL 37% hydrochloric acid (Merck), and five drops of hydrofluoric acid (HF). After etching, the chip was rinsed thoroughly in DI water for 10 minutes to remove residual chemicals. Subsequently, the chip was placed in an ultrasonic acetone bath to remove any residual nanoscale PZT, fully exposing the underlying Pt layer, with the process carefully monitored (Figure 3(c)). A 1.5 µm-thick silicon dioxide ($SiO_2$) layer was deposited over the entire chip by plasma-enhanced chemical vapor deposition (PECVD) to encapsulate and protect the Al layer for subsequent steps (Figure 3(d)). The chip was again spin-coated with AZ 5214E photoresist and patterned using the second mask in image reversal mode to create an inverse-tone pattern. A metal bilayer consisting of 30 nm chromium (Cr) and 330 nm aluminum (Al) was then deposited and lifted off to form a hard mask for etching the underlying layers (Figure 3(e)).

In the next step, the deposited SiO₂ passivation layer, along with the underlying Pt and Ti layers, as well as the SiO₂ on the device layer surface, were sequentially etched using a Sentech SI 500 ICP-RIE system. This process involved the successive application of SiO₂-etching and metal-etching recipes to expose the underlying silicon (Si) device layer. The SiO₂ layers were etched using a soft-etching mode monitored in real time with an interferometric endpoint detection system to ensure precise stop at the Si interface. Deep reactive ion etching (DRIE) of the silicon device layer was then carried out using the Oxford PlasmaLab System 100 ICP 300 Deep RIE tool. As the wafer initially contained a 50 µm-thick device layer, but the target structural thickness is 130 µm, etching proceeded through the device layer and continued by removing the buried oxide (BOX) layer using SiO₂-selective etching. Subsequently, deep silicon etching was resumed into the handle layer using DRIE until the total etch depth reached 130 µm (Figure 3(f)).

The third and final photolithography mask was applied to the backside of the chip. A chromium/aluminum (Cr/Al) bilayer was deposited and lifted off to define a hard mask for the DRIE process. The underlying SiO₂ layer was first etched to expose the silicon handle layer, which was then etched using DRIE to a depth of approximately 115 µm (Figure 3(h)). Subsequently, the Cr/Al and SiO₂ layers in the release regions were etched using appropriate etch recipes. Noting that during these etching steps, the substrate frame was protected by a simple thermal tape applied to the backside. This protection preserves the integrity of the backside hard mask on the substrate frame, while additional etching was performed in the release regions, including the D-D shaped support structures and the area beneath the outer frame, to define the backside pattern of the mirror. DRIE was continued until the movable structure was fully released, leaving the support regions with a remaining thickness of approximately 175 µm (Figure 3(i)).

As the final processing step, the 1 cm × 1 cm device chip is fully released following the removal of all layers, including hard masks and SiO₂. Using a 10 µl pipette, a tiny droplet of PZT etchant solution (from previous steps) is carefully applied to one corner of the chip to etch the PZT, exposing the bottom electrode pad (Figure 3(j)). Finally, the chip is washed thoroughly with DI water, acetone, and 2-propanol to yield the completed micro-scanner chip as shown in Figure 2.

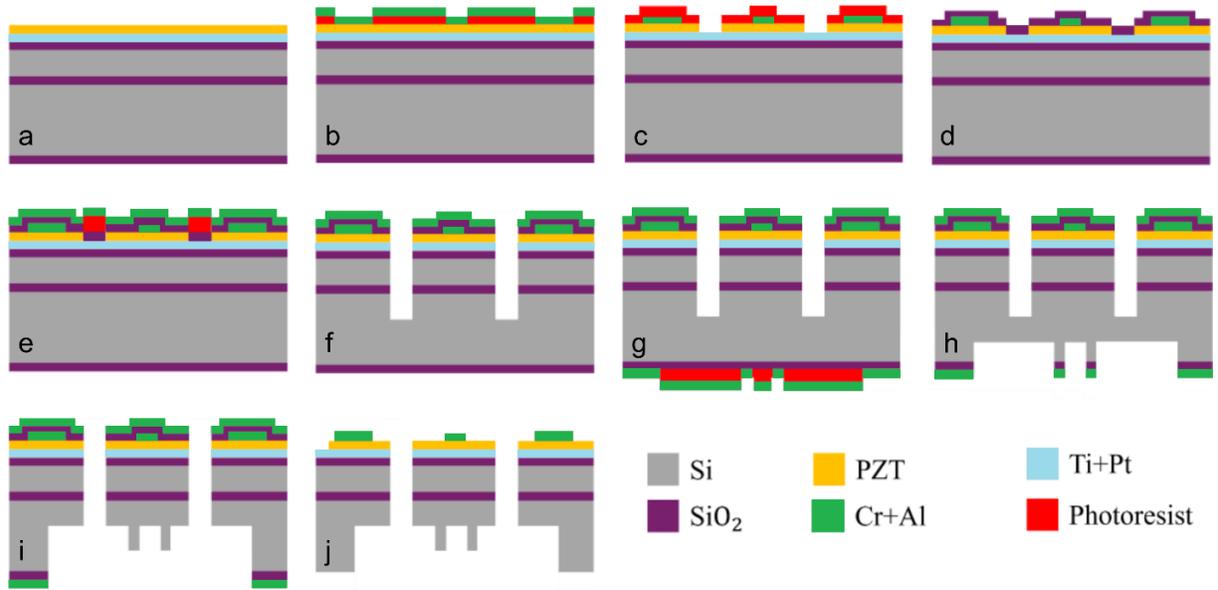

Figure 3. Fabrication process flowchart. (a) Initial wafer. (b) (Mask-1) Patterning of top electrodes, mirror, connections, and pads via lift-off. (c) (Mask-2) Etching of the PZT layer. (d) Deposition of SiO₂ to encapsulate and protect all features for the subsequent steps. (e) (Reverse Mask-2) Patterning of the hard mask. (f) RIE and DRIE etching on the top side. (g) (Mask-3) Patterning of the hard mask on the backside. (h) RIE and DRIE etching on the backside. (i) Final DRIE and release of the device structure. (j) Removal of the hard mask and SiO₂, and then opening of the bottom electrode pad.

### 3. Finite Element Analysis

The finite element analysis (FEA) of the piezoelectric scanner was performed using COMSOL Multiphysics software. The FEA was conducted based on the proposed scanner design, as described in the scanner architecture and design section, incorporating all relevant features of the device, including the overall dimensions and the selected SOI wafer thickness. Material properties were defined using COMSOL's built-in library, and the mesh was generated with high quality using the sweeping method, where an extremely fine mesh size was applied to ensure simulation accuracy.

Modal analysis was conducted to extract the eigenmodes of the scanner. As illustrated in Figure 5(a), the first mode occurs at 3.718 kHz, where the outer frame exhibits resonant behavior on its torsional beams.

This mode, characterized by clear torsional resonance, is well-suited for vertical scanning due to its ability to produce linear and stable motion along the vertical axis. The eighteenth mode, observed at 54.504 kHz, demonstrates out-of-phase motions between the inner and middle frames, as shown in Figure 5(b). This high-frequency mode is ideal for horizontal scanning, facilitating rapid, small-amplitude oscillations that enable precise horizontal movement. Figures 5(c) and 5(d) illustrate the stress distribution, which remains well below the conservative fracture strength of single-crystal silicon (typically reported around 1–2 GPa) [43]. Stress is primarily concentrated at the torsional beams and the folded beams connecting the PZT actuators to the outer frame during vertical scanning, as well as at the torsional beam junction between the middle and inner frames. Stress distribution analysis reveals that the stress levels remain significantly below the silicon stress limit, ensuring the mechanical integrity of the scanner under operational conditions. These patterns are consistent with previous MEMS design studies, where torsional beams are designed to handle high loads without exceeding material failure thresholds [42]. The results confirm that the structure will maintain reliable performance over extended periods of use.

Dynamic deformation of the mirror is a critical factor in determining the optical quality of the reflected beams in MEMS scanners. To minimize stress and deformation, the mirror is mechanically decoupled from the inner frame and supported by a stiffened D-D shaped rim on the backside. Additionally, the outer frame is reinforced with a stiffened rim to enhance stability. The dynamic deformation profiles of the outer frame and mirror are shown in Figures 5(c) and 5(d). The results confirm that the mirror surface remains suitable for diffraction-limited imaging, as most of the deformation occurs near the edges and is negligible for optical performance.

The frequency-domain simulation and mechanical response of the scanner structure are presented in Figure 5 (e-f). These simulations provide valuable insights into the dynamic behavior of the device, particularly its resonant frequencies. For the vertical scanning mode, the primary resonance peak is observed at 3.7185 kHz, aligning well with the frequency-domain simulation results (Figure 5(e)). Similarly, the horizontal scanning mode exhibits a resonance peak at 54.504 kHz (Figure 5(f)). The distinct and narrow displacement peaks at these frequencies reflect the mechanical responsiveness and structural precision of the device. These results suggest that the scanner design supports consistent resonant behavior in both vertical and horizontal directions, making it potentially suitable for high-resolution optical applications.

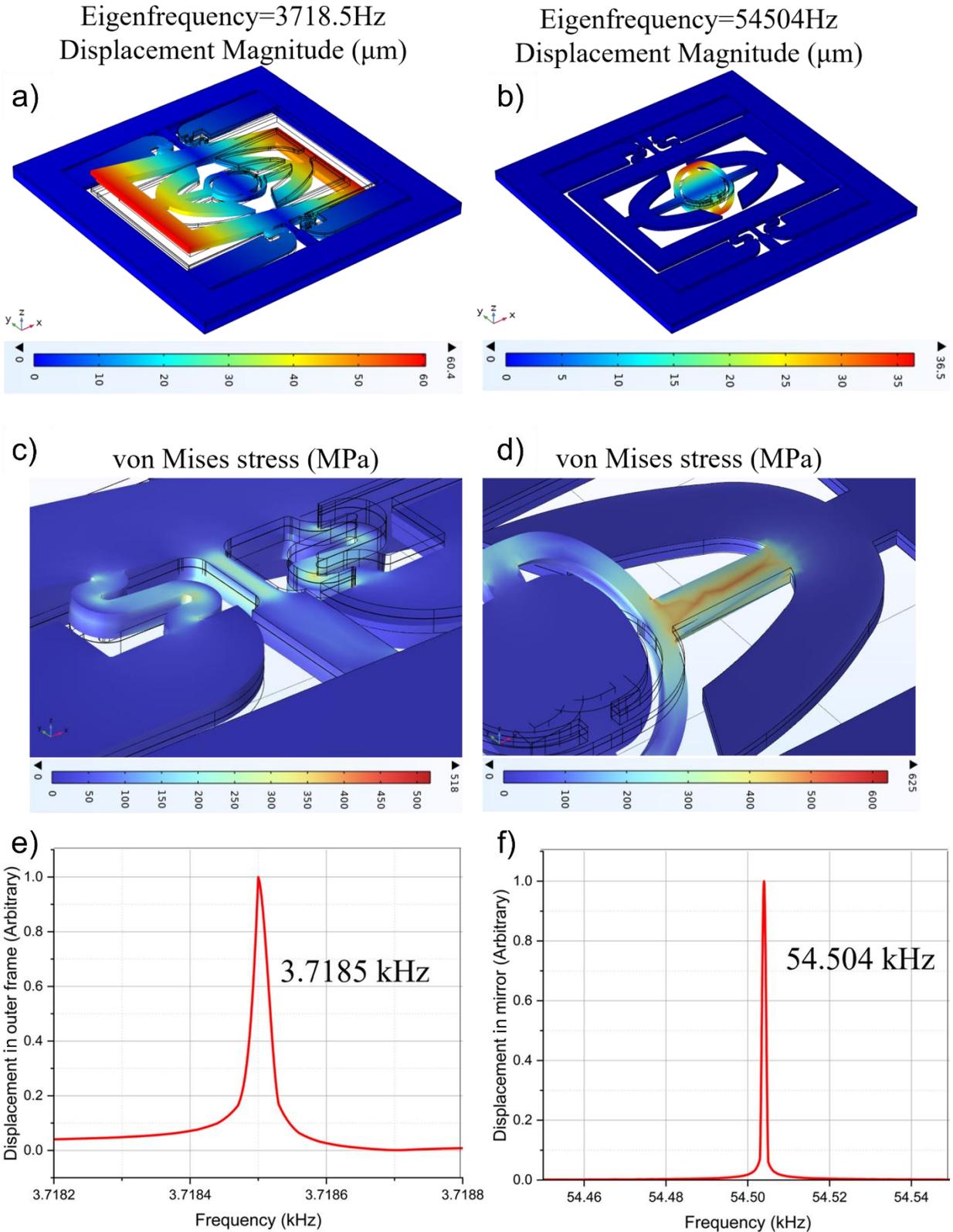

Figure 5. (a) Modal analysis of the first mode at 3.718 kHz, showing the outer frame resonating on its torsional beams, which is suitable for vertical scanning. (b) Modal analysis of the eighteenth mode at 54.504 kHz, illustrating the out-of-phase motions of the inner and middle frames, which is suitable for horizontal scanning. (c–d) Stress distribution analysis, with stress concentrated on the torsional and folded beams, confirms that it remains significantly below the silicon stress limit. (e) Frequency-domain simulation and mechanical response of the scanner structure, where the vertical scanning mode exhibits a maximum resonance peak at 3.718 kHz, and (f) the horizontal scanning mode shows a maximum peak at 54.504 kHz.

## 4. Characterization and Performance Results

Figures 6(a) and (b) illustrate the optical setup of the prototype fabricated PZT scanner. Two signal generators, each providing two outputs with a 180° phase difference, actuate the PZT scanner. One generator drives the vertical actuator pads, while the other drives the horizontal actuator pads. The grounds of both generators are interconnected, and then connected to the scanner's ground pad. A red laser serves as the illumination source, with the optical path outlined in Figure 6(a). The laser beam (represented by red arrows) is first directed by a beamsplitter, which divides it into reflected and transmitted components. The reflected beam illuminates the surface of the microscanner, while the transmitted beam continues along the optical axis. The microscanner redirects the beam, which is then divided once more by the beamsplitter into transmitted and reflected portions. The transmitted scanning beam is directed onto a mirror and projected onto a screen, as shown in Figure 6.

Activation of the vertical actuator pads enables one-dimensional (1D) vertical scanning, as illustrated in Figure 6(b). Similarly, horizontal 1D scanning is achieved by activating the horizontal actuator pads, as shown in Figure 6(c). Simultaneous actuation of both sets of pads results in two-dimensional (2D) scanning, as depicted in Figure 6(d), captured at a projection distance of 60 cm between the MEMS scanner and the screen. Figure 6(e) displays representative Lissajous patterns generated during 2D scanning, acquired at a scanning distance of 60 cm from the MEMS scanner to the screen. In this configuration, the horizontal and vertical optical scanning angles were calculated to be 11.5° and 4.8°, respectively. The applied voltage to the scanner during these operations is an AC periodic pulse 12 V peak-to-peak, with a frequency of 54.175 kHz for horizontal 1D scanning and 3.6 kHz for vertical 1D scanning.

Resonance frequency measurements were conducted using the Polytec MSA-600 Micro System Analyzer (Figure 7(a)). The MEMS scanner was bonded via double-sided tape to a piezo disk (Figure 7(b)) for mechanical coupling and actuated at a maximum voltage of 2 V, with a frequency sweep used to identify its natural resonance modes. Figures 7(c) and (d) display the measured vertical and horizontal resonance frequencies, along with their estimated corresponding quality factor (QF). Specifically, the vertical resonance frequency was measured at 3.6 kHz with a QF of 300.56, while the horizontal resonance frequency was measured at 54.175 kHz with a QF of 642.76. It was observed that the measured resonance frequencies were slightly lower than those predicted by simulations. Such deviations can typically be attributed to fabrication tolerances, e.g., DRIE process variations (resulting in an approximate 14 µm over-etch in some cases, Figure 2(b)), along with material property differences between simulation assumptions and fabricated device properties. These deviations can result in a reduction in effective mass or stiffness changes, thus lowering resonance frequencies. Overall, these observations align with previous literature on MEMS fabrication, where process variations affect the effective stiffness and mass of the structures, thereby changing the resonant frequencies.

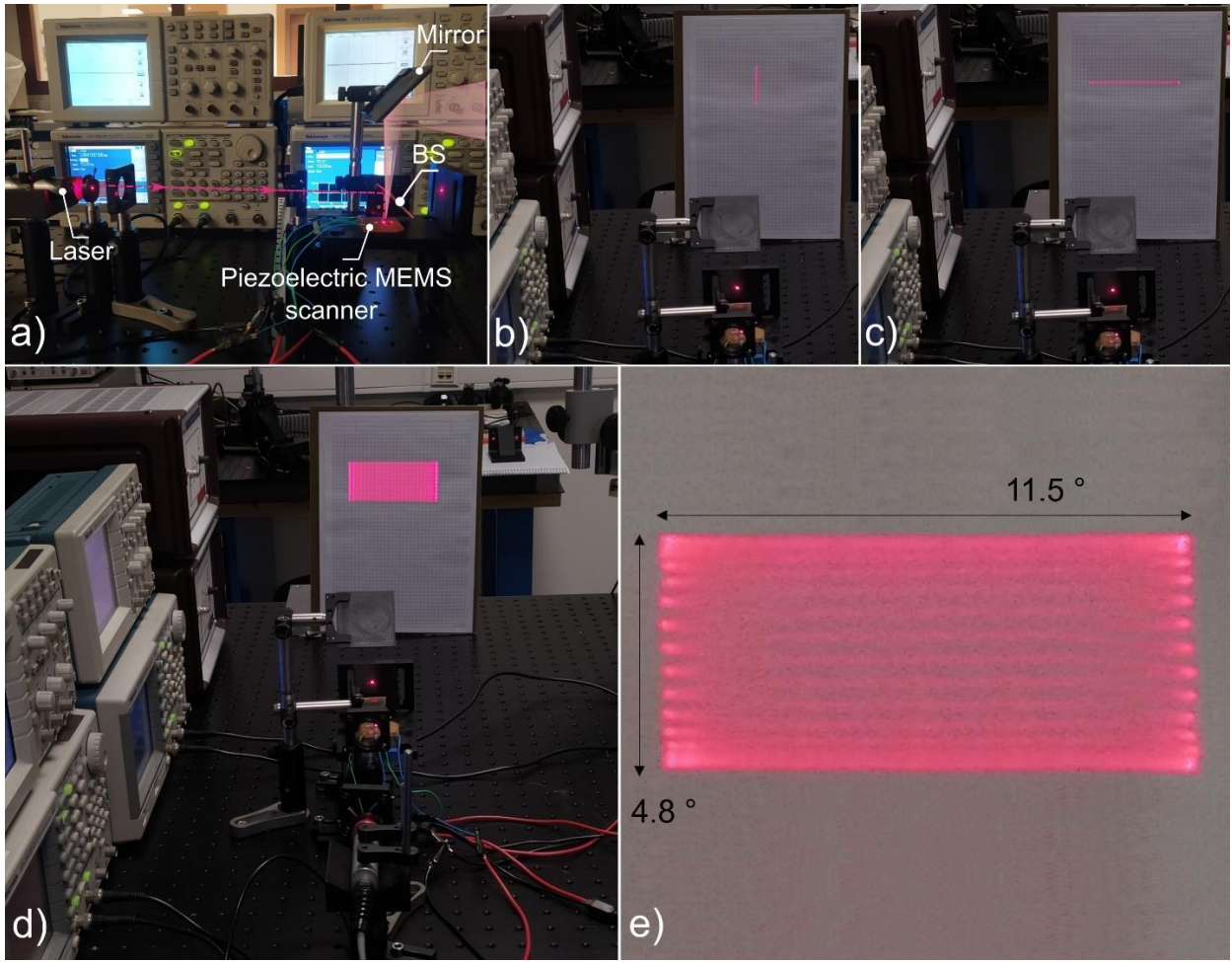

*Figure 6. Optical setup and scanning results of the MEMS scanner. (a) Experimental setup showing the optical path: the laser beam is directed by the beamsplitter onto the PZT scanner and reflected as a scanning beam projected onto the screen. (b) Vertical one-dimensional (1D) scanning achieved by activating only the vertical actuator pads. (c) Horizontal 1D scanning achieved by activating only the horizontal actuator pads. (d) Two-dimensional (2D) scanning obtained through simultaneous activation of both vertical and horizontal actuator pads. (e) Representative Lissajous patterns generated during 2D scanning at a projection distance of 60 cm from the screen, with a horizontal scanning angle of 11.5° and a vertical scanning angle of 4.8°. (BS: beamsplitter).*

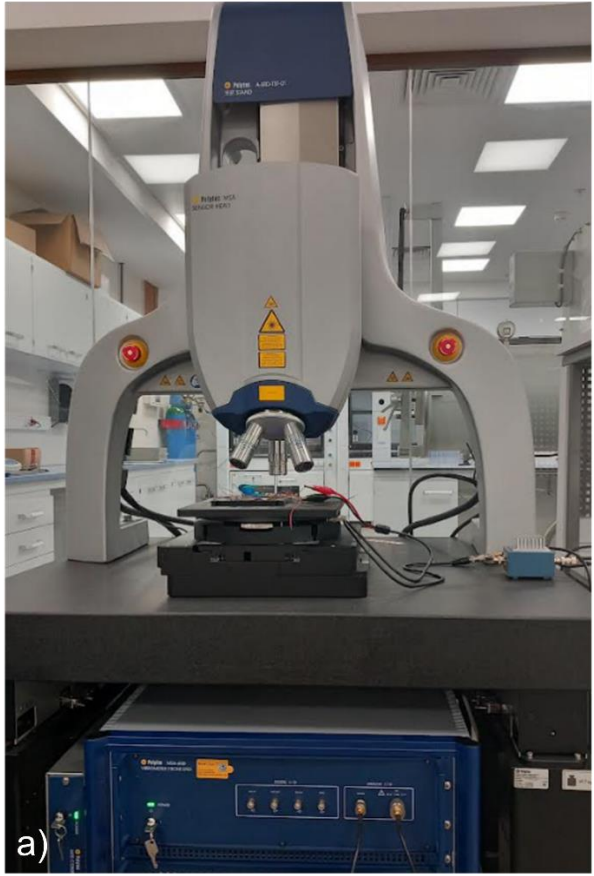
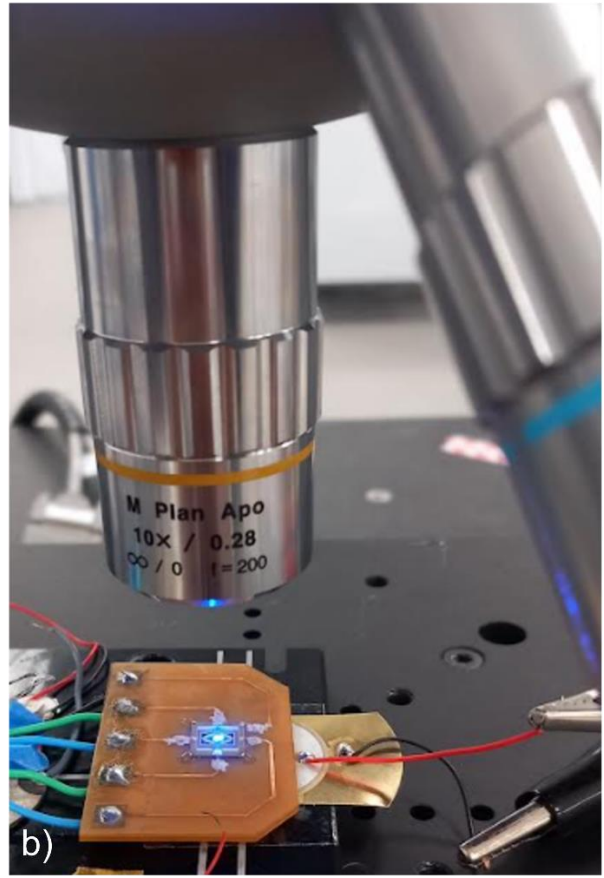
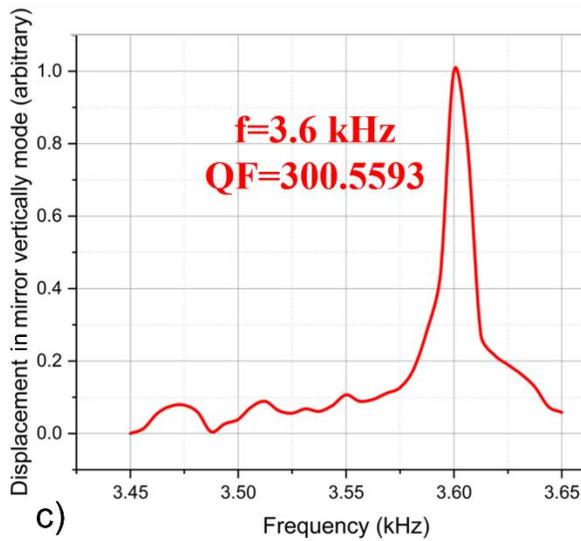
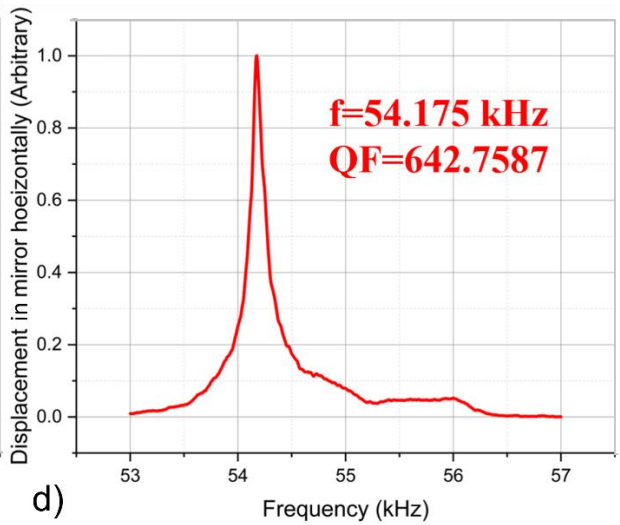

*Figure 7. Natural resonance frequency measurement setup and results. (a) Polytec MSA-600 Micro System Analyzer was used to measure the resonance frequencies. (b) MEMS scanner mounted on a piezo disk for actuation; the scanner was driven with a maximum applied voltage of 2 V, and a frequency sweep was performed to determine its natural resonance modes. (c) Measured vertical resonance frequency of 3.6 kHz with a QF of 300.56. (d) Measured horizontal resonance frequency of 54.175 kHz with a QF of 642.76.*

The voltage response presented in Figure 8 was obtained by systematically varying the actuation voltage while operating the MEMS scanner at its natural resonance frequency, measured in the optical setup as shown in Figure 6. This method ensures that at each voltage level, the maximum displacement response of the scanner is accurately captured. For each data point, the peak displacement was determined by measuring both the distance between the screen and the PZT scanner, and the width of the projected scanning beam, which directly correlates with the scanner's angular displacement and provides a reliable representation of its dynamic behavior. The experimental results illustrate the relationship between the applied voltage and the scanner's angular displacement, underscoring the efficiency of the piezoelectric actuation mechanism in converting electrical energy into mechanical motion. Notably, as observed, scan performance is stabilized at approximately 12 Vp-p; beyond this point, further increases in driving voltage yield diminishing returns in scan angle. This saturation can be attributed to the inherent mechanical and electromechanical limitations of piezoelectric materials, such as material stiffness and piezoelectric coupling efficiency. Figure 8(b) illustrates the acquired optical scan angle as a function of the applied frequency sweep for both the vertical and horizontal actuation modes separately, with an input voltage of 12 Vp-p. The observed fundamental resonances, which are in-line with the finite element simulation results, are at 3.6 kHz and 54.175 kHz for the vertical and horizontal modes, respectively. Based on these measurements, the QF of the designed PZT MEMS scanner is calculated to be 750 in vertical mode and 1050 in horizontal mode. Further, the optical scanning bandwidth efficiency product, given by the product of the optical scanning angle ($\theta\_opt$), effective mirror diameter (D), and resonant frequency (f), is calculated 24.2 deg·mm·kHz for the vertical mode and 623 deg·mm·kHz for the horizontal mode. The differences in these values reflect distinct operational frequencies and scanning amplitudes between the vertical and horizontal modes, which are consistent with performance metrics reported for similar MEMS scanning systems. A high bandwidth-efficiency product indicates that the scanner is capable of achieving wide scanning angles at high speeds, a critical requirement for applications such as high-resolution optical imaging, microscopy, and laser projection systems. Overall, this comprehensive characterization provides valuable insights into the operational limitations and optimal driving conditions of the piezoelectric MEMS scanner, enabling informed design decisions and enhanced performance in practical applications [4, 22].

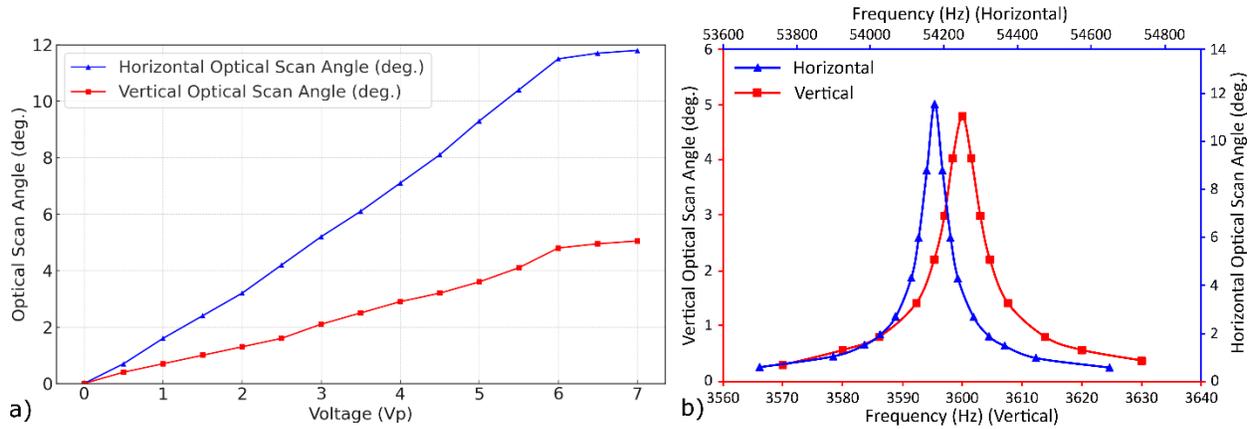

*Figure 8. a) Measure the scanning angle by applying actuation voltage at 3.6 kHz vertically and 54.175 kHz horizontally. b) The QF of the proposed MEMS scanner is measured to be 750 and 1050 for the vertical and horizontal modes, respectively.*

## 5. Discussion

The dynamic behavior of the fabricated PZT-actuated MEMS scanner was characterized through two experimental setups: mechanical resonance analysis using the Polytec MSA-600 Micro System Analyzer (Figure 7a and b) and the optical response evaluation within the optical scanning setup (Figure 6a). Each measurement provided an insight into the scanner's performance, especially in terms of the mechanical-natural resonance frequency, which is the same, and the QF, which is distinct.

Resonance frequency measurements in the Polytec MSA-600 Micro System Analyzer, were performed by bonding the MEMS scanner to a piezo-disc using double-sided tape for mechanical coupling and actuating it with a maximum input voltage of 2 V. The scanner exhibited fundamental resonances at 3.6 kHz (vertical) and 54.175 kHz (horizontal) -same as the optical setup measurement setup- while corresponding QFs of 300.56 and 642.76, respectively as shown in Figure 6. The relatively lower QFs observed in this setup can be attributed to two primary factors: (1) the bonding interface for mechanical coupling, which introduces mechanical damping and energy loss due to the compliant nature of the adhesive layer, and (2) the limited excitation voltage, which restricts the amplitude of vibration and the precision of frequency response measurement. In contrast, the optical measurements conducted under full resonance conditions (Fig. 8) yielded significantly higher QFs of 750 and 1050 for the vertical and horizontal modes, respectively. These measurements were performed by operating the scanner at its natural resonance frequencies (measured in the Polytec MSA-600 Micro System Analyzer) and the actuation voltage at 12 Vp-p. This setup allowed direct observation of the scanning beam's angular displacement, enabling more accurate assessment of the mechanical response. The improved QFs in this case are indicative of reduced energy losses in the optical setup and stronger mechanical excitation, highlighting the impact of system integration and actuation conditions on resonance behavior. Importantly, as shown in Figure 8a, measurements of the scanning angle by applying actuation voltage at 3.6 kHz vertically and 54.175 kHz horizontally in the optical scanning setup show the presence of actuation saturation effects. The scan angle increased with voltage up to approximately 12 Vp-p, and beyond which further increases yielded negligible improvement in angular displacement. This saturation behavior can arise from inherent electromechanical limitations in piezoelectric actuators, including polarization saturation, mechanical damping, and nonlinear strain response under high electric fields, as well as our proposed scanner architecture [35, 44-46]. These findings align with previous studies that have identified flexure stiffening, increased dielectric losses, and hysteresis as key contributors to performance plateauing in piezoelectric-MEMS systems [35, 45].

In addition, the QF and overall scanning efficiency may be further influenced by the microstructural composition of the MEMS device. The presence of a buried oxide (BOX) layer in SOI-based MEMS platforms is known to introduce acoustic impedance mismatches and internal friction, which can significantly degrade QF performance [47]. Some studies show that the acoustic impedance mismatch between $SiO_2$ and adjacent materials, like silicon, can lead to reflections and scattering of acoustic waves, further contributing to energy loss. However, when carefully engineered, studies have shown that optimizing the oxide interface and optimizing the acoustic properties of $SiO_2$ can enhance the QF of MEMS resonators [48, 49].

To quantify system performance, the optical scanning bandwidth-efficiency product (θ_opt·D·f) was calculated as 24.2 deg·mm·kHz in vertical and 623 deg·mm·kHz in horizontal, demonstrating the scanner's suitability for applications, such as Imaging, Microscopy, Optical Coherence Tomography (OCT), and Ophthalmology applications, requiring high-speed angular deflection with compact footprints [22, 50-52]. Overall, the study, including the design, fabrication, and characterization, demonstrated the feasibility of the proposed 2D PZT MEMS scanner architecture as a proof of concept, showing its capability to operate effectively in two-dimensional scanning mode.

### 6. Conclusion and Future Work

In this work, we have presented the design, fabrication, and characterization of a novel architectural 2D MEMS scanner actuated by PZT thin films, with its potential for applications such as ophthalmological imaging, laser scanning, and optical coherence tomography (OCT). Furthermore, the study successfully represented the proof of concept for the proposed 2D PZT MEMS scanner architecture, achieved through an efficient three-mask fabrication process, highlighting its feasibility and efficiency for practical applications. The scanner's structural design and frequency response were thoroughly evaluated, with simulated and measured resonance frequencies showing good agreement. The measured resonant frequencies of 3.6 kHz for the vertical mode and 54.2 kHz for the horizontal mode, along with QFs of 750 and 1050, respectively, validate the system's robust performance. Notably, these QFs, especially in the horizontal operation, indicate effective energy sustainability, which is significant for stable scanning and coupled with a large scanning angle at resonance. The optical characterization further highlighted the scanner's capability to perform 1D and 2D scanning, including Lissajous patterns with scan angles of 11.5° (horizontally) and 4.8° (vertically) with an AC pulse signal of peak-to-peak amplitude of 12 V. The demonstrated optical scanning bandwidth-efficiency product performance, 24.2 (vertical) deg·mm·kHz and 623 (horizontal) deg·mm·kHz, places it among the higher 2D piezoelectric MEMS scanners reported. The proposed MEMS scanner combines mechanical stability with sufficient scanning angles and high-actuation precision, making it a promising candidate for imaging applications such as OCT. Its integration with OCT systems would have the ability to substantially enhance resolution and image depth for industrial and biomedical applications. Overall, this study offers potential design parameters for the improvement of the optical performance of a 2D piezoelectric scanning mirror.

The future work can be focused on minimizing the structural flaws by optimizing the fabrication process, employing a SOI wafer of device layer thickness 150 or 170 µm, modifying the folded beam of the scanner, and integrating the scanner designed further into applications (like head up display, OCT, Lidar, pico-projector) to optical components for their compatibility. Furthermore, the incorporation of adaptive control systems like piezoelectric sensing elements and small-scale electronics might even enhance the device's precision and functionality for practical applications.

## Acknowledgment

This study was supported by the TÜBİTAK 2247 program (support agreement number 120C145) and the European Innovation Council (EIC) Transition program (support agreement number 101057672). The authors acknowledge the resources and support provided by the N2STAR cleanroom core facilities at Koç University and the SUNUM cleanroom core facilities at Sabancı University. The authors would like to thank Sven Holmstrom for his feedback and advice on fabrication, Çağla Selalmaz from N2STAR, and Süleyman Çelik from the SUNUM cleanrooms for their assistance with device fabrication.